\documentclass[prl,reprint, twocolumn,showpacs,preprintnumbers,amsmath,amssymb,nofootinbib,floatfix]{revtex4-1} 
\usepackage{graphicx}

\usepackage{mathrsfs}
\usepackage{hyperref}

\usepackage{slashed}

\usepackage{color}

\def \matrix #1 {\left(\begin{array}{cc} #1 \end{array}\right)}

\def \e  {\mathop{\rm e}\nolimits}

\def\II{\hbox{{1}\kern-.25em\hbox{l}}}
\def \Li{\mathop{\rm Li}\nolimits}

\newcommand \widebar [1] {\overline{#1}}

\newcommand \vev [1] {\langle{#1}\rangle}

\newcommand \ket [1] {|{#1}\rangle}
\newcommand \bra [1] {\langle {#1}|}

\begin{document}

\title{Next-to-Next-to-Leading-Order QCD Prediction for the Photon-Pion Form Factor}

\author{Jing Gao}
\email{gaojing@ihep.ac.cn}
\affiliation{School of Physics, Nankai University, \\ Weijin Road 94, 300071 Tianjin, China}
\affiliation{Naturwissenschaftlich-Technische Fakult\"at, Universit\"at Siegen, \\ Walter-Flex-Str.~3, 57068 Siegen, Germany}
\affiliation{Institute of High Energy Physics, CAS, P.O. Box 918(4) Beijing 100049, China}
\affiliation{School of Physics, University of Chinese Academy of Sciences, Beijing 100049, China}
\author{Tobias Huber}
\email{corresponding author: huber@physik.uni-siegen.de}
\affiliation{Naturwissenschaftlich-Technische Fakult\"at, Universit\"at Siegen, \\ Walter-Flex-Str.~3, 57068 Siegen, Germany}
\author{Yao Ji}
\email{corresponding author: yao.ji@uni-siegen.de}
\affiliation{Naturwissenschaftlich-Technische Fakult\"at, Universit\"at Siegen, \\ Walter-Flex-Str.~3, 57068 Siegen, Germany}
\author{Yu-Ming Wang}
\email{corresponding author: wangyuming@nankai.edu.cn}
\affiliation{School of Physics, Nankai University, \\ Weijin Road 94, 300071 Tianjin, China}

\date{\today}

\begin{abstract}
\noindent
We accomplish the complete two-loop computation of the leading-twist contribution
to the photon-pion transition form factor $\gamma \, \gamma^{\ast} \to \pi^0$
by applying the hard-collinear factorization theorem together with modern multi-loop techniques.
The resulting predictions for the form factor
indicate that the two-loop perturbative correction
is numerically important.
We also demonstrate that our results will play a key role
in disentangling various models of the twist-two pion distribution amplitude
thanks to the envisaged precision at Belle II.\\[0.4em]

\end{abstract}

\preprint{SI-HEP-2021-16, P3H-21-036}

\maketitle

%
\section{introduction}
%

It is widely accepted that the exclusive two-photon production of a light
pseudoscalar meson $\pi^0$  is of utmost importance for probing the partonic landscape
of composite hadrons with unprecedented precision
and for deepening our understanding of the factorization properties of
hard exclusive QCD reactions in general.
Historically, the photon-pion transition form factor (TFF) had been explored intensively
(see, for instance~\cite{Cornwall:1966zz,Gross:1972dd,Brodsky:1971ud})
even before the advent of QCD due to its apparent connection to the axial anomaly
\cite{Adler:1969gk,Bell:1969ts,Bardeen:1969md}
when taking the vanishing virtuality limits for the two photons.
Likewise, for the energetic photo-production at the light-like distance,
the dynamical behavior of the TFF
$\gamma^{\ast} \, \gamma^{(\ast)} \to \pi^0$
can be predicted by applying the operator-product-expansion (OPE) technique for
the time-ordered product of two  electromagnetic currents \cite{Lepage:1979zb,Lepage:1980fj}.
The appearing correlation function with the distinct kinematic set-up
concurrently provides the theory description of a variety of two-photon processes including
deeply inelastic lepton-hadron scattering (DIS) \cite{Feynman:1969wa,Bjorken:1968dy,Bjorken:1969ja}
and deeply virtual Compton scattering (DVCS) \cite{Muller:1994ses,Ji:1996nm,Radyushkin:1996nd},
whose  importance in establishing QCD as the theory of strong interactions
and in accessing the transverse-distance structure of the target hadron
cannot be exaggerated (see, for instance \cite{Gross:1975vu,Collins:2011zzd,Dudek:2012vr,Accardi:2012qut}).
In particular, the double-virtual photon-to-pion form factor has been demonstrated to be
an indispensable ingredient  for determining the hadronic light-by-light scattering (HLbL) contribution
to  the anomalous magnetic moment of the muon
in the dispersive framework~\cite{Colangelo:2014dfa}.
The comprehensive data-driven strategy
of determining the pion-pole contribution  to the HLBL scattering further illustrates that
the yielding hadronic uncertainties of $(g-2)_{\mu}$ can be reduced substantially
with the improved determination of the single-virtual photon-pion form factor
\cite{Hoferichter:2018dmo,Hoferichter:2018kwz}.
Moreover, the collinear QCD dynamics dictating the TFF at large momentum transfer
is encapsulated in the universal light-cone distribution amplitude (DA),
which is undoubtedly of decisive importance
for the model-independent description of semileptonic and nonleptonic $B$-meson decays,
such as $B \to \pi \, \ell^{-} \, \bar \nu_{\ell}$, $B \to \pi \, \ell^{+} \, \ell^{-}$
and $B \to \pi \, \pi$,   on the basis of QCD factorization
demanding the detailed information on the very pion DA
as the fundamental non-perturbative input \cite{Beneke:2000wa,Beneke:2003pa,Beneke:2004dp,Beneke:2001ev}.
Additionally, the  lattice QCD technique has been  applied to address
the photon-pion form factor at low photon virtualities  with very encouraging predictions
\cite{Gerardin:2016cqj,Gerardin:2019vio}.

Experimentally, the pion TFF $\gamma \, \gamma^{\ast} \to \pi^0$
with one on-shell and one off-shell photon turns out to be more accessible
from the ``single-tagged" measurements of the differential
$e^{+} \, e^{-} \to e^{+} \, e^{-} \, \pi^{0}$ cross section~\cite{Gronberg:1997fj,Aubert:2009mc,Uehara:2012ag}.
The unexpected scaling violation of the TFF
reported by the BaBar measurements~\cite{Aubert:2009mc}
triggered a storm of enthusiasm in the theory community with diverse speculations
from both perturbative and non-perturbative QCD perspectives.
The subsequent Belle data~\cite{Uehara:2012ag} covering the same kinematical region
$Q^2\in[4 , 40] \, {\rm GeV^2}$ (where $Q^2$ stands for the virtuality of the off-shell photon),
however, did not reveal  the pronounced increase of the pion TFF
in the high-$Q^2$ region thus creating a moderate tension at the level of
$2 \sigma$ in comparison with the BaBar results.
The substantial improvement on the integrated luminosity
and the trigger efficiency of the  Belle II experiment at SuperKEKB~\cite{Kou:2018nap}
will evidently enable a clarification of
the observed discrepancy between the BaBar and the Belle data
and urgently demand the high-precision theory computation
of the $\gamma \, \gamma^{\ast} \to \pi^0$ TFF in QCD.

At leading power (LP) in $\Lambda_{\rm QCD}^2 / Q^2$ the  photon-pion TFF
can be expressed in terms of the perturbatively calculable 
coefficient function (CF) and the twist-two pion DA~\cite{Lepage:1980fj}
(see also \cite{Li:1992nu,Musatov:1997pu,Li:2013xna}).
The short-distance CF at the next-to-leading order (NLO)
in $\alpha_s$ had been determined
more than three decades ago~\cite{delAguila:1981nk,Braaten:1982yp,Kadantseva:1985kb}.
The next-to-next-to-leading-order (NNLO) computation of the perturbative matching coefficient
was further carried out in the large-$\beta_0$ approximation~\cite{Melic:2001wb}
and in the so-called conformal scheme ~\cite{Melic:2002ij}.
Consequently, the factorization-scale invariance of the resulting expression
for the pion TFF was not achieved in the modified minimal subtraction ($\widebar{\rm MS}$) scheme at two loops.
Accomplishing the full NNLO QCD prediction to the TFF $\gamma \, \gamma^{\ast} \to \pi^0$
is therefore of the top priority, on the one hand,  for developing the perturbative factorization formalism
of hard exclusive reactions;
and on the other hand, for elevating our capability
to confront the yielding theory predictions
with the forthcoming precision of Belle II measurements.
In this Letter, we will take advantage of the QCD
factorization program to obtain analytically the two-loop hard function
by  evaluating an appropriate bare QCD matrix element at ${\cal O}(\alpha_s^2)$
by means of dedicated multi-loop computational strategies
and by implementing  the ultraviolet (UV) renormalization and
the infrared (IR) subtractions,
with an exploratory phenomenological analysis regarding the numerical significance
of the predicted TFF.

%
\section{The photon-pion form factor}
%

We first set up the theory framework for establishing the hard-collinear factorization formula
of the TFF $F_{\gamma \pi} (Q^2)$. The latter is defined in terms of the matrix element of the electromagnetic current
\begin{equation}
j_{\mu}^{\rm em} = \sum_{q} \, g_{\rm em} \, e_q \, \bar q \, \gamma_{\mu} \, q
\end{equation}
between an on-shell photon with momentum $p^{\prime}$ and a pion with momentum $p$
\begin{align}
\langle \pi(p) | j_{\mu}^{\rm em}  | \gamma (p^{\prime}) \rangle
= g_{\rm em}^2 \epsilon_{\mu \nu \alpha \beta} q^{\alpha}  p^{\beta}  \epsilon^{\nu}(p^{\prime})
F_{\gamma \pi} (Q^2) \,,
\end{align}
where $q=p-p^{\prime}$ refers to the transfer momentum,
$\epsilon^{\nu}(p^{\prime})$ is the polarization vector of the on-shell photon,
and $e_q$ denotes the electric charge of the quark field in units of $g_{\rm em}$
(the positron charge).
Moreover, we have employed the convention $\epsilon_{0123}= -1$ and the notation $Q^2=-q^2$.
Introducing further two light-cone vectors $n_{\mu}$ and $\bar n_{\mu}$
with  $n^2=\bar n^2=0$ and $n \cdot \bar n=2$
allows for the decomposition $p_{\mu} = (\bar n \cdot p) /2 \, n_{\mu}$
and $p^{\prime}_{\mu} = (n \cdot p) /2 \, \bar n_{\mu}$.

Applying the hard-collinear factorization theorem results in the LP contribution
to the $\gamma \, \gamma^{\ast} \to \pi^0$  form factor
\begin{align}
F^{\rm LP}_{\gamma\pi}(Q^2)&=\frac{(e_u^2-e_d^2)f_\pi}{\sqrt{2}\,Q^2}
\int^1_0dx\,T_2(x,Q^2,\mu_F)\,\phi_\pi(x,\mu_F) \nonumber \\[0.3em]
&\equiv \, \frac{(e_u^2-e_d^2)f_\pi}{\sqrt{2}\,Q^2}
T_2(x,Q^2,\mu_F)\otimes \phi_\pi(x,\mu_F) \,,
\label{eq:factorization}
\end{align}
where the pion decay constant $f_\pi=(130.50  \pm 0.02 \pm 0.03 \pm 0.13)~{\rm MeV}$ is determined from
the experimental measurement \cite{Zyla:2020zbs}
and $\mu_F$ represents the factorization scale for which we tacitly adopt
$\mu_F=\mu_{\rm UV}\equiv \mu$ in \eqref{eq:factorization}.
Here the $\pi^0$ wavefunction $\ket{\pi^0}=\frac{1}{\sqrt{2}}(\ket{u\bar u}-\ket{d\bar d})$
with exact isospin symmetry is employed.
The CF $T_2$ can be expanded perturbatively in the form (similarly for any other partonic quantity)
\begin{align}
T_2=\sum_{\ell=0}^\infty  a_s^{\ell} \, T_2^{(\ell)} \,,
\qquad a_s\equiv\frac{\alpha_s}{4\pi}  \,.
\label{eq:loopexpansion}
\end{align}
The twist-two pion DA $\phi_\pi(x,\mu_F)$ is defined by the renormalized  matrix element
on the light-cone
\begin{align}
&\bra{\pi(p)}\,[\bar q(z\bar n)[z\bar n,0]\gamma_\mu\gamma_5 q(0)]_R\,\ket{0}
\nonumber \\
&=-i f_\pi p_\mu\int^1_0dx\,\e^{i x z \bar n \cdot p} \, \phi_\pi(x,\mu_F)\,,
\label{def:pion-LCDA}
\end{align}
where $[z\bar n,0]$ is the Wilson line to ensure gauge invariance.
We also note that $\phi_\pi(x,\mu_F)=\phi_\pi(\bar x,\mu_F)$ due to charge symmetry, where $\bar x\equiv 1- x$.
The  corresponding renormalization group (RG) equation at one loop ~\cite{Efremov:1979qk,Lepage:1980fj}
implies the series expansion of  $\phi_\pi(x,\mu_F)$ in terms of Gegenbauer polynomials,
\begin{align}
\phi_\pi(x,\mu_F) =  6 \, x \, \bar x \, \sum_{n=0}^{\infty}
a_{n}(\mu_F) \, C_{n}^{3/2}(2 \, x -1)\,,
\label{Gegenbauer expansion of pion-LCDA}
\end{align}
with the vanishing odd-moments $a_{1, 3, ...}(\mu_F)$.

It is convenient to evaluate the
CF
by inspecting the following correlation function
\begin{align}
    &\frac{g_{\rm em}^2 e_q^2}{2\,\bar n\cdot p}\,\Pi_{\mu\nu} = i\int \!\! d^4z \e^{-iq\cdot z}\nonumber \\
    &\qquad\qquad
    \times \bra{\bar q(\bar xp) q(x p)}
    {\rm T} \{ j^{\rm em}_\mu(z), \, j^{\rm em}_\nu(0) \}\ket{0} \,,
\end{align}
which can be parameterized by the two form factors for the bilinear quark currents
with the spin structures \cite{Wang:2017ijn}
\begin{align}
 \Gamma_{A}^{\mu\nu} &= \gamma_{\perp}^{\mu} \, \slashed{\bar n} \, \gamma_{\perp}^{\nu} \;, &
 \Gamma_{B}^{\mu\nu} &= \gamma_{\perp}^{\nu} \, \slashed{\bar n} \,  \gamma_{\perp}^{\mu}  \;,
\end{align}
thanks to the QED Ward identity and charge symmetry. The perpendicular component of any four-vector $p^\mu$ is defined by $p^\mu = (n p) \bar n^\mu/2 + (\bar n p) n^\mu/2 + p^\mu_\perp$.
We will dedicate the next section to the two-loop calculation of the bare quantity $\Pi_{\mu\nu}^{(2)}$
and subsequently derive the master formula for the CF $T_2$ in \eqref{eq:factorization}.
%

%
\section{Two-loop calculation}
%

\begin{figure}[htp]
\includegraphics[width=0.21 \textwidth]{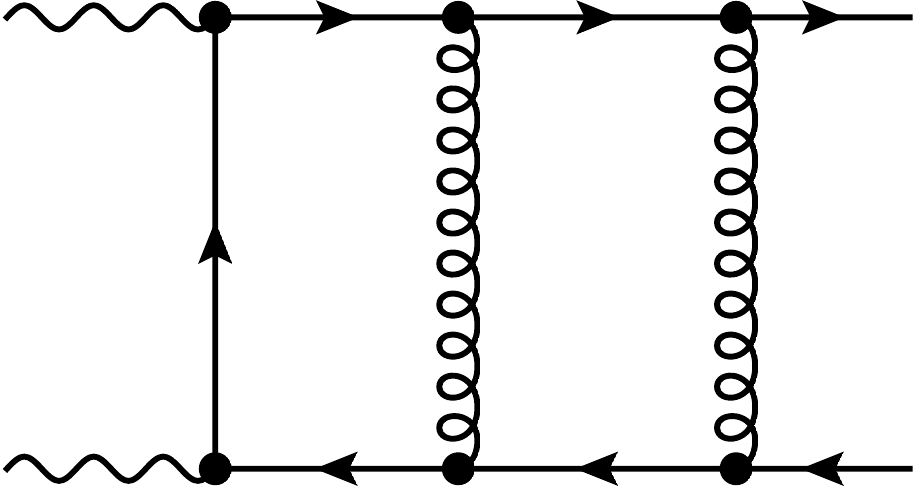}\hspace*{10pt}
\qquad
\includegraphics[width=0.21  \textwidth]{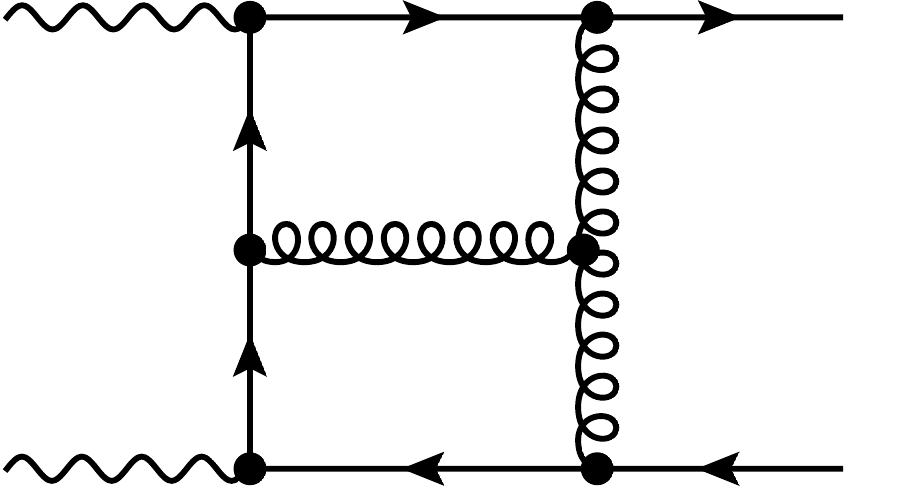}\hspace*{10pt}
\caption{Sample Feynman diagrams. }
\label{fig:diagrams}
\end{figure}

The techniques entering the calculation of the bare two-loop amplitude
have become standard in contemporary multi-loop computations.
We first generate the Feynman diagrams with both {\tt Feynarts}~\cite{Hahn:2000kx}
and by means of an in-house routine. Sample diagrams are shown in Fig.~\ref{fig:diagrams}.
After taking into account that certain diagrams have color factor zero,
vanish due to the Furry theorem and/or represent the flavor-singlet contributions,
a set of $42$ diagrams (plus the ones with the two photons exchanged) has to be computed.
The entire calculation is carried out in dimensional regularization with $d=4-2\epsilon$.
%

The Dirac and tensor reductions are performed with in-house routines, yielding two-loop scalar integrals,
which get further processed with {\tt FIRE}~\cite{Smirnov:2008iw}, an implementation of
Laporta's algorithm~\cite{Laporta:1996mq,Laporta:2001dd} relying on integration-by-parts identities~\cite{Chetyrkin:1981qh,Tkachov:1981wb}.
In addition, we exploit the fact that for the quark momenta $p_1 \equiv x p \propto \bar x p \equiv p_2$, which yields relations between integrals based on momentum conservation and
enables us to arrive at the minimal set of master integrals.
At this stage, additional Dirac structures besides $\Gamma_{A,B}^{\mu\nu}$
disappear from the sum of all diagrams, making the QED Ward identity and charge symmetry manifest and providing a valuable check of our calculation.

In total, we get $12$ independent master integrals,
the  analytic $\epsilon$-expansion of which can be written in terms of harmonic polylogarithms (HPLs)~\cite{Remiddi:1999ew,Maitre:2005uu,Maitre:2007kp}.
We relegate the explicit analytic results of all master integrals
to a forthcoming longer write-up.

%
\section{Ultraviolet Renormalization and Infrared Subtractions}
%

We are now prepared to extract the two-loop contribution to $T_2$ by first
introducing the operator basis $\{ O_{1}^{\mu \nu}, \, O_{2}^{\mu \nu}, \, O_{E}^{\mu \nu} \}$  with
\begin{align}
O_{j}^{\mu \nu}(x)= \frac{\bar n \cdot p}{2 \pi} \int d \tau  \,
 \e^{{ i \bar x  \tau \bar n \cdot p}}
\bar q (\tau \bar n) \, [\tau \bar n,0]\,  \Gamma_{j}^{\mu \nu}  \,  q(0) \,,
\end{align}
 where
\begin{align}
\Gamma_{1}^{\mu \nu} &= g_{\perp}^{\mu \nu} \slashed{\bar n}  \,, \quad
\Gamma_{2}^{\mu \nu} =  i \epsilon_{\perp}^{\mu \nu} \slashed{\bar n}  \gamma_5 \,,\quad
\Gamma_{E}^{\mu \nu} =
 \slashed{\bar n}\sigma_\perp^{\mu\nu}
-  \Gamma_{2}^{\mu \nu} \,,
\end{align}
with $\sigma_\perp^{\mu\nu}=\frac12[ \gamma_{\perp}^{\mu},  \gamma_{\perp}^{\nu}]$ and then by exploiting the matching equation
\begin{align}
\Pi^{\mu\nu} = \sum\limits_{a=1,2,E} T_a \otimes \langle O_{a}^{\mu\nu} \rangle\, .
\label{eq:matching}
\end{align}
It is evident from the definitions  $g_{\perp}^{\mu \nu} = g^{\mu \nu}-n^{\mu} \bar n^{\nu}/2 -n^{\nu} \bar n^{\mu}$/2
and $\epsilon^{\mu \nu}_{\perp}  \equiv \epsilon^{\mu \nu \alpha \beta} \bar n_{\alpha} \, n_{\beta}/2$
that $O_{E}^{\mu \nu}$ is an evanescent operator vanishing at $d=4$,
and the CP-even operator $O_{1}^{\mu \nu}$ cannot couple to the pseudoscalar $\pi^{0}$ state.
We therefore encounter a unique physical operator $O_{2}^{\mu \nu}$~\cite{Wang:2017ijn}.

The correlator $\Pi^{\mu\nu}$ assumes the following form to all orders in $\alpha_s$
in terms of the tree-level matrix elements of the effective operators $O_{1, 2,  E}^{\mu \nu}$
\begin{align}
    \Pi^{\mu\nu}= \!\!\sum_{k=1,2,E}\, \sum_{\ell=0}^\infty  (Z_\alpha  a_s)^\ell\,  A_k^{(\ell)}(x)\,\vev{\bar q(\bar xp)\Gamma_{k}^{\mu\nu}q(xp)}^{(0)}
    \, , \label{eq:correlator}
\end{align}
where
$Z_\alpha = 1 - a_s \beta_0/\epsilon + {\cal O}(a_s^2)$.
Note that, hereafter, we disregard $O_1^{\mu\nu}$ completely
due to parity. The expansion of $T_2$ was already given in Eq.~(\ref{eq:loopexpansion}),
and due to scaleless integrals in dimensional regularization
the matrix elements of the light-cone operators are expanded as
\begin{align}
\vev{O_a^{\mu\nu}}
= & \sum_{\ell=0}^\infty a_s^\ell \, Z_{ab}^{(\ell)} \otimes \vev{O_b^{\mu\nu}}^{(0)}  \label{eq:MEOa}\\
 = &  \left\{ \delta_{ab} + a_s \, Z_{ab}^{(1)} +  a_s^2 \, Z_{ab}^{(2)} + {\cal O}(a_s^3)\right\} \otimes \langle O_b^{\mu\nu} \rangle^{(0)} \, . \nonumber
\end{align}
Sums over repeated indices are understood to run over $\{2,E\}$, $\langle O_{a}^{\mu\nu} \rangle^{(0)}$ is the tree-level matrix element of $O_a^{\mu\nu}$,
and the renormalization constants $Z_{22}^{(\ell)}$ can be extracted from the
Efremov-Radyushkin-Brodsky-Lepage (ERBL) kernel with
the one- and two-loop result given in~\cite{Efremov:1979qk, Lepage:1980fj} and~\cite{Sarmadi:1982yg,Dittes:1983dy,Katz:1984gf,Mikhailov:1984ii,Belitsky:1999gu}, respectively.
Inserting Eqs.~\eqref{eq:loopexpansion},~\eqref{eq:correlator}, and~\eqref{eq:MEOa} into Eq.~\eqref{eq:matching} and comparing coefficients of $\langle O_{2,E}^{\mu\nu} \rangle^{(0)}$ at ${\cal O} (a_s^2)$ leads to the following ``master formula'' for the CF at NNLO
\begin{align}
T_2^{(2)} &= A_2^{(2)} + Z_\alpha^{(1)} A_2^{(1)} - \sum_{a=2,E}\sum_{k=0}^1 Z_{a2}^{(2-k)}\otimes T_a^{(k)}  \, .
\label{eq:masterformula}
\end{align}
All quantities multiplied by a divergent term  in~\eqref{eq:masterformula}
must be expanded beyond ${\cal O}(\epsilon^0)$ to correctly capture all finite terms.
It remains important to emphasize the role of the evanescent operator in the master formula~\eqref{eq:masterformula}. Despite the vanishing of evanescent operators in four dimensions, their non-vanishing bare matrix element and evanescent-to-physical mixing (captured by $Z_{E2}$~\cite{Dugan:1990df})
 are crucial for correctly implementing the IR subtraction and establishing the hard-collinear factorization in our $d$-dimensional framework. These steps are necessary in view of the fact that 
 taking the limit $d \to 4$ does not commute
with the hard-collinear factorization for the correlation function $\Pi^{\mu\nu}$
(see also \cite{Braun:2003wx,Beneke:2009ek}).

Substituting the newly derived two-loop correction to the hard function,
whose explicit expression is given in the supplemental material,
into the factorization formula (\ref{eq:factorization}) and adopting further the asymptotic pion DA
enable us to write down
\begin{widetext}
\begin{align}
F^{\rm LP,\,  Asy}_{\gamma\pi}(Q^2) & = \frac{(e_u^2-e_d^2)f_\pi}{\sqrt{2} \, Q^2} \,
\bigg\{ 6 - 30 \, a_s \, C_F - a_s^2 \, \bigg[ C_F \beta_0 \, \left( \left (31+12 \log\frac{\mu^2}{Q^2} \right )\zeta_2
+ 6 \zeta_3 + 7 \right )
\label{asymtotic factorization formula}
\\
& \quad \left.\left.+ \, C_F^2 \, \left(24(\zeta_2+\zeta_3) \log\frac{\mu^2}{Q^2}
+42\zeta_4+54\zeta_3+37\zeta_2-\frac{85}{2} \right)
-\frac{C_F}{N_c} \, \left(6\zeta_4-12\zeta_3-2\zeta_2+13 \right ) \right ]
+ {\cal O} (\alpha_s^3) \right \} \,.\notag
\end{align}
\end{widetext}
Generalizing (\ref{asymtotic factorization formula}) to account for  the non-asymptotic correction
will give rise to lengthy expressions due to the intricacy of the NNLO CF.
Consequently, we will only  evaluate such higher moment
effect numerically in the phenomenological exploration.
In particular, it has been verified explicitly  that the photon-pion form factor computed from
the factorization formula (\ref{eq:factorization}) is  independent of the factorization scale at ${\cal O}(\alpha_s^2)$,
with the aid of the two-loop evolution equation of the twist-two pion DA ~\cite{Sarmadi:1982yg,Dittes:1983dy,Katz:1984gf,Mikhailov:1984ii,Belitsky:1999gu}.
Subsequently, the complete next-to-next-to-leading-logarithmic (NNLL) resummation
of $\log (Q^2 / \Lambda_{\rm QCD}^2)$ appearing in~\eqref{eq:factorization} is achieved
 by virtue of the three-loop evolution equation of $\phi_\pi(x,\mu_F)$
in the naive dimensional regularization (NDR) scheme \cite{Braun:2017cih},
which is implemented throughout our computation.

%
\section{Numerical analysis}
%

We now turn to explore the  numerical  implications
of the achieved NNLL computation of 
$F_{\gamma \pi} (Q^2)$
with three phenomenological models of $\phi_{\pi}(x, \mu_0)$ at $\mu_0 = 1 \, {\rm GeV}$,
\begin{eqnarray}
\allowdisplaybreaks
{\rm Model \,\, I:} \,\,\, && \phi_{\pi}(x, \mu_0) =
\frac{\Gamma(2 + 2 \, \alpha_{\pi})}{  \Gamma^2(1 + \alpha_{\pi})}  \,
(x \, \bar x)^{\alpha_{\pi}} \,,
\label{Models of the pion DA} \\
&& {\rm with } \,\, \alpha_{\pi}(\mu_0) = 0.422^{+0.076}_{-0.067}  \,;
\qquad
\nonumber \\[0.6em]
{\rm Model \,\, II:} \,\,\, && \{a_{2},a_{4}\}(\mu_0) = \{0.269(47), 0.185(62)\} \,,
\nonumber \\
&& \{a_{6}, a_8\}(\mu_0) = \{ 0.141(96), 0.049(116) \} \, ;
\nonumber \\[0.6em]
{\rm Model \,\, III:} \,\,\, && \{ a_{2},a_{4} \}(\mu_0) = \{0.203^{+0.069}_{-0.057} \,,
-0.143^{+0.094}_{-0.087}\} \,.
\nonumber
\end{eqnarray}
Model I is inspired from the AdS/QCD correspondence~\cite{Brodsky:2007hb}
with modifications from the most recent lattice result $a_{2}(2 \, {\rm GeV})=0.116^{+0.019}_{-0.020}$~\cite{Bali:2019dqc},
which corresponds to $a_{2}(\mu_0)=0.176^{+0.027}_{-0.029}$.
The construction of Model II
is achieved by matching the light-cone QCD sum rules of the pion electromagnetic form factor
in the space-like region onto the modified dispersion integral of
the modulus of the time-like form factor \cite{Cheng:2020vwr}.
The intervals of $a_2$ and $a_4$
in Model III~\cite{Bakulev:2001pa,Mikhailov:2016klg,Stefanis:2020rnd}
are determined from the method of QCD sum rules
~\cite{Mikhailov:1991pt}.
To facilitate the implementation of the three-loop  evolution of
the leading-twist pion DA \cite{Braun:2017cih,Strohmaier:2018tjo},
we will apply the Gegenbauer expansion
for the ``holographic"-type Model I by discarding $a_{2 n}(\mu_0)$ with $n \geq 7$
(see also \cite{Agaev:2010aq}).

%
%
%
\begin{figure}[htp]
\begin{center}
\includegraphics[width=0.8 \columnwidth]{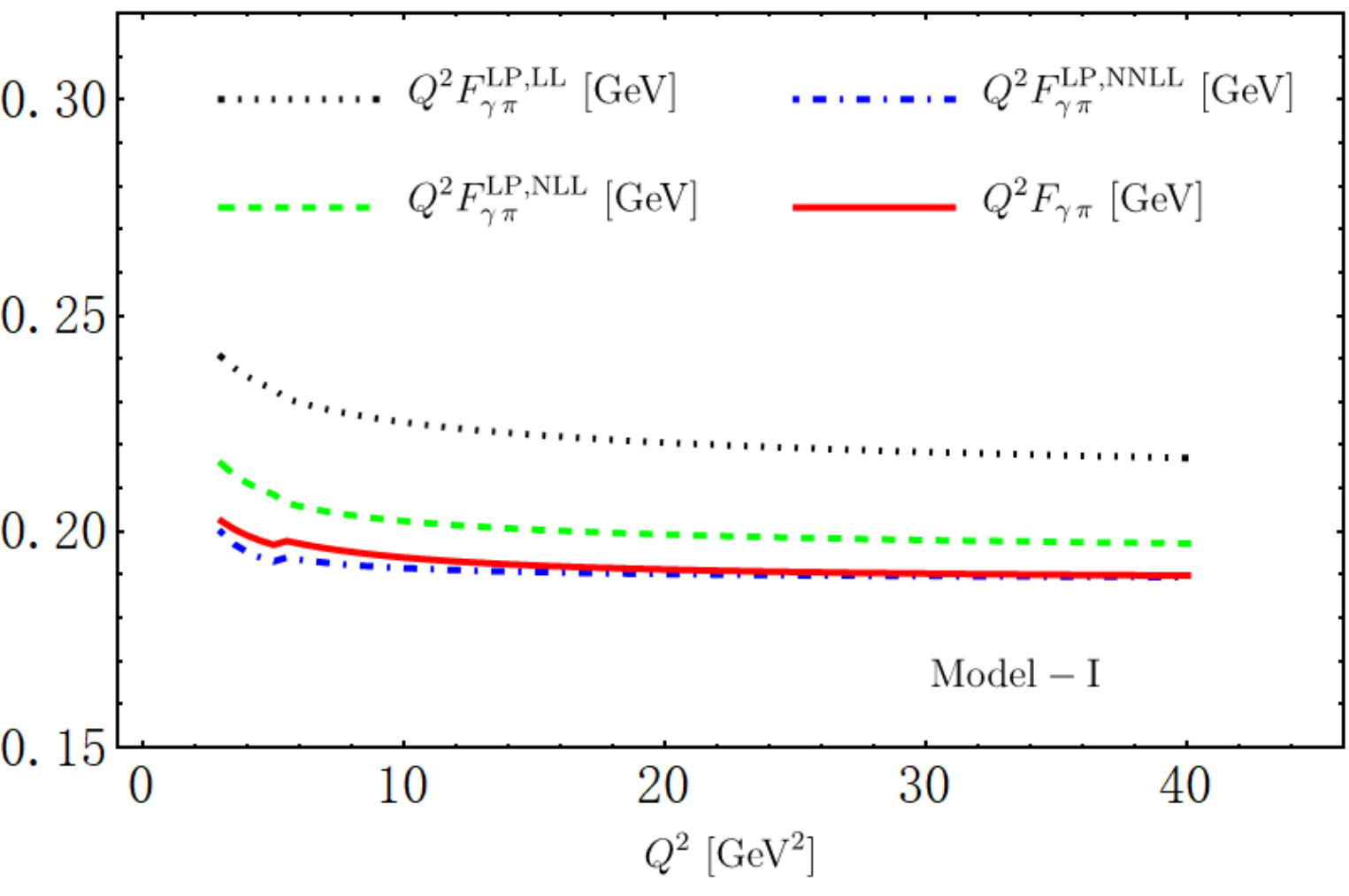}
\vspace*{0.1cm}
\caption{The twist-2 predictions of $F_{\gamma\pi}$
at the leading-logarithmic (LL) order [black dotted],
at the next-to-leading-logarithmic (NLL) order [green dashed],
and at the NNLL order [blue dot-dashed]
in QCD for Model I given in~\eqref{Models of the pion DA},
where the red solid curve is obtained by adding further
the subleading power contributions evaluated in \cite{Khodjamirian:1997tk,Wang:2017ijn}.}
\label{fig: Breakdown of the pion FF for Model-I}
\end{center}
\end{figure}
%
%
%

Adopting Model I as our default choice,
we present in Fig.~\ref{fig: Breakdown of the pion FF for Model-I}
the resulting impacts of QCD radiative corrections to the TFF at LP
as well as the numerical features of the power suppressed terms
from the twist-four collinear dynamics of the pion system  \cite{Khodjamirian:1997tk}
and from the long-distance photon correction \cite{Wang:2017ijn}.
Inspecting the numerical patterns of the various dynamical mechanisms
displayed in Fig.~\ref{fig: Breakdown of the pion FF for Model-I} implies
that the NNLL twist-two correction can reduce the corresponding
NLL prediction by an amount of $(4 \sim 7) \, \%$
at $Q^2\in[3 , 40]~ {\rm GeV^2}$
(in agreement with the previous estimate \cite{Mikhailov:2016klg}),
while the genuine one-loop perturbative correction
is responsible for a constant $\sim10 \%$ reduction
of the LL contribution in the same kinematic domain.
By contrast,
the above-mentioned subleading contributions are observed to bring about the negligible corrections~\cite{Wang:2017ijn}.
The  full two-loop computation of the LP effect in the TFF is therefore indeed of extraordinary phenomenological significance.
Additionally, the three-loop QCD correction estimated with the ``na\"{\i}ve non-Abelianization" prescription \cite{Beneke:1994qe}
is expected to be  approximately ${\cal O} (10-20)  \%$ of the one-loop radiative effect
without adding further the systematic uncertainty.
%

%
%
%
\begin{figure}[htp]
\begin{center}
\includegraphics[width=0.8 \columnwidth]{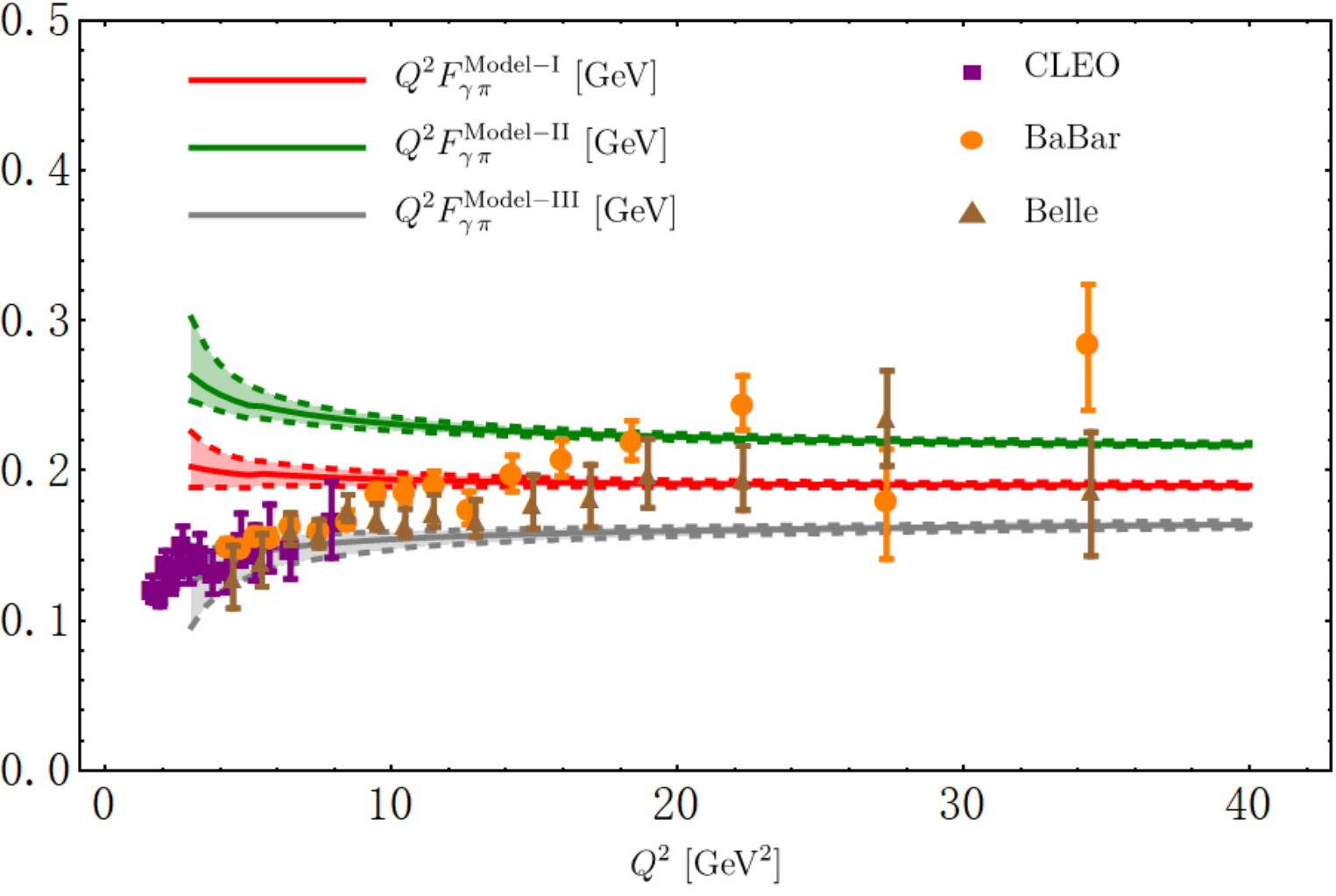}
\vspace*{0.1cm}
\caption{Theory predictions of the TFF
$\gamma\gamma^{\ast} \to \pi^0$ for the three models presented
in~\eqref{Models of the pion DA}.
The color bands are due to variation of the factorization/renormalization scale $\mu$.
For a comparison, we also display the experimental measurements
from the CLEO~\cite{Gronberg:1997fj} (purple squares),
BaBar~\cite{Aubert:2009mc} (orange circles)
and Belle \cite{Uehara:2012ag} (brown spades) Collaborations.}
\label{fig: Model dependence of the pion FF}
\end{center}
\end{figure}

We proceed to present in Fig.~\ref{fig: Model dependence of the pion FF}
the theory predictions of the TFF for the three phenomenological models of the leading-twist pion DA.
As the dominating non-perturbative uncertainties of our
numerical analysis from $\phi_{\pi}(x, \mu_0)$ can be inferred from
the discrepancies of the obtained theoretical results for the distinct sample models,
the quoted uncertainties in \eqref{Models of the pion DA} are
not included in the three shaded (red, green, gray) bands
in Fig.~\ref{fig: Model dependence of the pion FF}
to highlight the perturbative uncertainty
from the variation of $\mu^2 = \langle x \rangle \, Q^2$
with $1/4 \leq \langle x \rangle \leq 3/4$~\cite{Agaev:2010aq}.
The distinctive snapshot of the well-separated uncertainty bands for the three models
is particularly encouraging to differentiate numerous speculations on the intricate behaviors
of the twist-two pion DA.
Furthermore,  the theory predictions from the incomplete two-loop computation of $F_{\gamma\pi}(Q^2)$
in the large $\beta_0$ approximation  differ from
our full ${\cal O} (\alpha_s^2)$ results by almost a factor of two at high-$Q^2$.
Additionally, the NNLL improved hard-collinear factorization
tends to validate the scaling behaviour of the photon-pion form factor at $Q^2 \geq 20 \, {\rm GeV^2}$
with an impressive precision of ${\cal O} (1 \, \%)$,
which will be confronted with the future Belle II data.
As expected, such scaling behaviour is  broken down
at low momentum transfer, numerically $Q^2 \leq 5 \, {\rm GeV^2}$,
due to the pronounced subleading power contributions.
A more detailed investigation of the asymptotic behaviour of $F_{\gamma\pi}(Q^2)$
in the hard-collinear factorization framework can be pushed forward
by introducing the scaling-rate quantity $\Omega(Q^2)$ as proposed in \cite{Stefanis:2019cfn,Stefanis:2020rnd}.

\begin{figure}[tp]
\begin{center}
\includegraphics[width=0.8 \columnwidth]{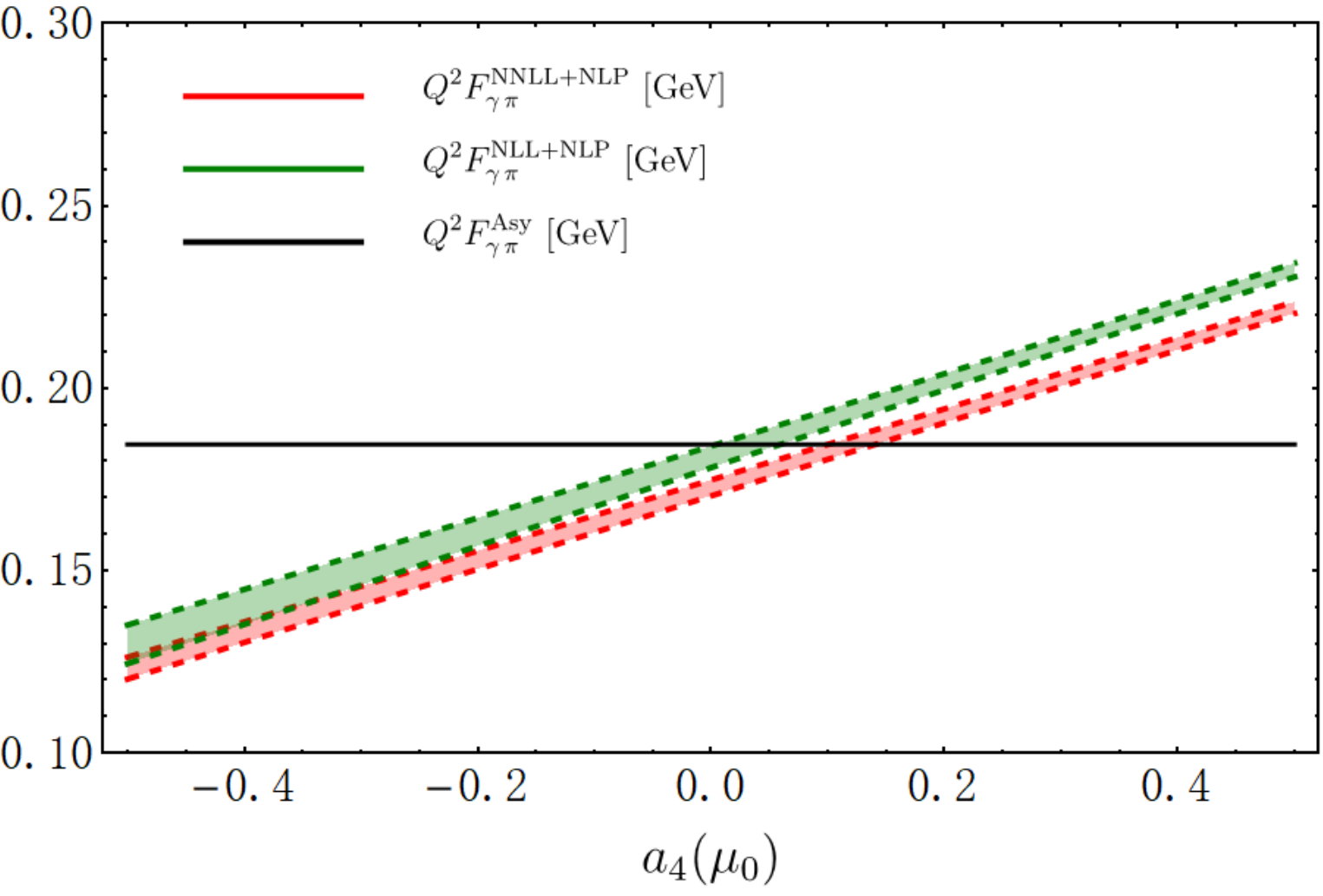}
\vspace*{0.1cm}
\caption{Illustration of the $a_4(\mu_0)$ dependence of the photon-pion form factor
at $Q^2=27.30 \, {\rm GeV^2}$ with the second Gegenbauer moment determined from~\cite{Bali:2019dqc}
and with the vanishing higher Gegenbauer moments
(i.e., $a_{n\geq 6}(\mu_0)=0$).}
\label{fig: a4-dependence of the pion FF}
\end{center}
\end{figure}

We finally address the actual impact of the NNLL correction
to the photon-pion form factor on the model-independent determination of the twist-two pion DA
in anticipation of the upcoming precision Belle II measurements.
To this end, we investigate the intrinsic sensitivity of the $\gamma \, \gamma^{\ast} \to \pi^0$
form factor with regard to the poorly constrained Gegenbauer moment $a_4(\mu_0)$,
by employing the improved lattice result for $a_2(\mu_0)$~\cite{Bali:2019dqc}
and by further leaving out the higher moments $a_{n\geq 6}(\mu_0)$  for illustration purpose.
The numerical implication displayed in Fig. \ref{fig: a4-dependence of the pion FF}
reveals that the  two-loop computation of the photon-pion TFF is indeed
highly beneficial for pinning down the theory uncertainty of the extracted value
of $a_4(\mu_0)$ (up to a factor of two improvement
approximately when compared with the corresponding one-loop result).

%
\section{Conclusions}
%

In summary, we have endeavored to carry out, for the first time, the complete two-loop computation
of the leading-twist contribution to the photon-pion form factor in the hard-collinear
factorization framework. The phenomenological significance of such  higher-order
perturbative effect was further addressed by employing the three non-perturbative models
of the twist-two pion DA.
The genuine ${\cal O}(\alpha_s^2)$ correction to the hard matching coefficient
was demonstrated to be numerically important at $Q^2\in [3,40]\, {\rm GeV^2}$,
and in particular, the previously evaluated radiative correction
at ${\cal O}(\alpha_s^2 \, \beta_0)$
only accounts for a moderate portion (approximately ${\cal O} (50 \, \%)$) of the full NNLL QCD effect.
Moreover, our improved theory predictions of the form factor $\gamma \, \gamma^{\ast} \to \pi^0$
appear to be promising for better constraining the shape parameters of the pion DA.
The established two-loop QCD factorization formula
of the photon-pion form factor is then naturally expected to be of notable importance
for exploiting  the many facets of the strong interaction dynamics
embedded in hard exclusive reactions.

%
\begin{acknowledgments}
\section*{Acknowledgements}
We would like to thank Ulrich Haisch for useful correspondence,
and Vladimir Braun and Alexander Manashov for sharing their results~\cite{Braun:2021grd}
with us prior to publication. We have found full agreement with their expression for the two-loop coefficient function.
Y.J. is also grateful to Alexander Manashov and Goutam Das for illuminating discussions.
J.G.  is partially supported by the Deutscher Akademischer Austauschdienst (DAAD).
The research of T.H. and Y.J. was supported in part by the Deutsche Forschungsgemeinschaft
(DFG, German Research Foundation) under grant  396021762 - TRR 257.
Y.M.W. acknowledges support from the National Youth Thousand Talents Program,
the Youth Hundred Academic Leaders Program of Nankai University,
the  National Natural Science Foundation of China  with Grant No. 11675082, 11735010 and 12075125,
and  the Natural Science Foundation of Tianjin with Grant No. 19JCJQJC61100.
\end{acknowledgments}
%

\bibliographystyle{apsrev4-1}

\bibliography{References}

\pagebreak
\widetext
\begin{center}
\textbf{\large Supplemental Materials to Next-to-Next-to-Leading-Order QCD Prediction for the Photon-Pion Form Factor}
\end{center}
\setcounter{equation}{0}
\setcounter{figure}{0}
\setcounter{table}{0}
\setcounter{page}{1}
\makeatletter
\renewcommand{\theequation}{S\arabic{equation}}
\renewcommand{\thefigure}{S\arabic{figure}}
\renewcommand{\bibnumfmt}[1]{[S#1]}
\renewcommand{\citenumfont}[1]{S#1}

%
\appendix

\begin{widetext}
\section{Analytic expressions in the \texorpdfstring{$\widebar{\rm MS}$}{Lg}-scheme}
%

The master formula~(14) was derived under the assumption that dimensional regularization is used on both sides of Eq.~(11) for both UV and IR divergences.
However, to determine the UV-renormalization constants $Z_{ab}^{(\ell)}$ we apply the following procedure (see e.g.~\cite{Beneke:2009ek,Wang:2017ijn}).
The renormalized matrix elements of the effective operators  are expressed as follows
\begin{align}
 \langle O_{a}^{\mu\nu} \rangle  =  \, \left\{ \delta_{ab} + a_s \, \left[M_{ab}^{(1)} + Z_{ab}^{(1)}\right] +  a_s^2 \, \left[M_{ab}^{(2)} + Z_{ab}^{(2)} \right.\right. 
 \left. \left. + Z_{a2}^{(1)} \otimes M_{2b}^{(1)}  \right] + {\cal O}(a_s^3)\right\} \otimes \langle O_b^{\mu\nu} \rangle^{(0)} \, ,
\end{align}
and the matrix elements $M_{ab}^{(\ell)}$ are obtained with dimensional regularization
applied only to the UV divergences  but with the IR regularization scheme being different from the dimensional one.
Renormalizing the matrix elements of the evanescent operator to zero \cite{Dugan:1990df}
yields the relations
\begin{align}
Z_{E2}^{(1)} & = - M_{E2}^{(1)} \, , &
Z_{E2}^{(2)} & = - M_{E2}^{(2)} + M_{E2}^{(1)} \otimes M_{22}^{(1)} \, ,
\end{align}
where $Z_{E2}^{(2)}$ is IR finite as expected albeit both $M_{E2}^{(2)}$ and $M_{22}^{(1)}$ are IR divergent. 

%
%
We now present the explicit expressions of the two-loop hard coefficient function $T_2^{(2)}$
with the NDR scheme of $\gamma_5$
%
\begin{align}
T_2^{(2)}(x) = & \, \beta_0 \, C_F \left( {\cal K}_{\beta}^{(2)}(x) / x + {\cal K}_{\beta}^{(2)}(\bar x) / \bar x \right)
             + C_F^2 \left({\cal K}_{F}^{(2)}(x) / x + {\cal K}_{F}^{(2)}(\bar x) / \bar x \right)
	     + C_F / N_c \left({\cal K}_{N}^{(2)}(x) / x + {\cal K}_{N}^{(2)}(\bar x) / \bar x \right) \, ,\label{eq:T22-expr}
\end{align}
where
{\allowdisplaybreaks
\begin{align}
{\cal K}_{\beta}^{(2)}(x) = & - L^2 \left(H_{0}(x)+\frac{3}{2}\right)
                              + L \left(-\frac{10}{3} H_{0}(x)-H_{1}(x)+2 H_{0,0}(x)-2 H_{1,0}(x)-2 \zeta_2-\frac{19}{2}\right)
			      -\zeta_2 H_{1}(x)-\frac{19}{9} H_{0}(x)\nonumber \\
                              &-\frac{1}{2} H_{1}(x)+\frac{10}{3} H_{0,0}(x)-\frac{14}{3} H_{1,0}(x)-H_{1,1}(x)-2 H_{0,0,0}(x)
			      +2 H_{1,0,0}(x)-H_{1,1,0}(x)-\frac{14}{3}  \zeta_2-\zeta_3-\frac{457}{24} \, ,\nonumber \\
                 \label{eq:K_beta}  & \\[1.0em]
{\cal K}_{F}^{(2)}(x) = & L^2 \left(6 H_{0}(x)-2 H_{1}(x)+4 H_{0,0}(x)+2 H_{1,0}(x)+\frac{9}{2}\right)
                        + L \left(8 \zeta_2 H_{0}(x)+\frac{38}{3} H_{0}(x)+4 \zeta_2 H_{1}(x)-17 H_{1}(x) \right. \nonumber \\
                        &\left. -6 H_{0,0}(x)+8 H_{1,0}(x)-2 H_{1,1}(x)-12 H_{0,0,0}(x)+4 H_{1,1,0}(x)-4 \zeta_3+6 \zeta_2+\frac{47}{2}\right)
			+ 6 \zeta_2 H_{0}(x)+4 \zeta_2 H_{1}(x) \nonumber \\
			&-2 \zeta_2 H_{2}(x)-8 \zeta_2 H_{0,0}(x)-2 \zeta_2 H_{1,0}(x)+2 \zeta_2 H_{1,1}(x)
			+32 \zeta_3 H_{0}(x)-4 \zeta_3 H_{1}(x)-\frac{64}{9} H_{0}(x)-\frac{71}{2} H_{1}(x) \nonumber \\
			&-\frac{38}{3} H_{0,0}(x)+\frac{34}{3} H_{1,0}(x)-11 H_{1,1}(x)
			-2 H_{1,2}(x)-8 H_{1,0,0}(x)+4 H_{1,1,0}(x)-2 H_{1,1,1}(x) \nonumber \\
			&-2 H_{1,2,0}(x)-4 H_{2,0,0}(x)-2 H_{2,1,0}(x)+12 H_{0,0,0,0}(x)-2 H_{1,0,0,0}(x)-2 H_{1,1,0,0}(x)\nonumber \\
			&+2 H_{1,1,1,0}(x)+3 \zeta_2^2+\frac{34}{3} \, \zeta_2 +39 \zeta_3+\frac{701}{24} \,,
            \label{eq:K_F}  \\[1.0em]
{\cal K}_{N}^{(2)}(x) = & L \left(4 \zeta_2 H_{0}(x)-\frac{8}{3} H_{0}(x)-4 H_{1}(x)-4 H_{3}(x)+4 H_{2,0}(x)+12 \zeta_3-1\right) + 12 \, x \left(\zeta_2 H_{0}(x)- H_{3}(x)+ H_{2,0}(x)\right) \nonumber \\
                        & -6 \zeta_2 H_{1}(x)-4 \zeta_2 H_{2}(x)-4 \zeta_2 H_{0,0}(x)+4 \zeta_2 H_{1,0}(x)+2 \zeta_2 H_{1,1}(x)
                          +14 \zeta_3 H_{0}(x)-\frac{32}{9} H_{0}(x)+11 H_{1}(x) \nonumber \\
                        &+4 H_{2}(x)+8 H_{4}(x)+\frac{8}{3} H_{0,0}(x)+\frac{2}{3} H_{1,0}(x)
			 +2 H_{1,1}(x)+6 H_{1,2}(x)-2 H_{1,3}(x)+6 H_{2,2}(x)-4 H_{3,0}(x) \nonumber \\
                        &-4 H_{3,1}(x)-6 H_{1,1,0}(x)-2 H_{1,1,2}(x)+4 H_{1,2,0}(x)-6 H_{2,0,0}(x)-4 H_{2,1,0}(x)-2 H_{1,1,0,0}(x) \nonumber \\
                        &+2 H_{1,1,1,0}(x)+\frac{1}{5} \, \zeta_2^2-\frac{22}{3} \,  \zeta_2+54 \zeta_3-\frac{73}{12} \,,
                        \label{eq:K_N}
\end{align}
}
with $L= \log (\mu^2/Q^2)$. The harmonic polylogarithms (HPLs) are defined recursively by
\begin{align}
H_{\vec0_n}(x)  &= \frac{1}{n!} \log^n (x) \,, &
H_{a_1 , a_2 , \ldots, a_n}(x)  &= \int_0^x dt \; f_{a_1}(t) \, H_{a_2, \ldots, a_n}(t)  \,.
\label{definition of HPLs}
\end{align}
\end{widetext}
The subscript of $H$ is called the weight vector. For the weight functions we have $f_{0}(t) = 1/t$, $f_{1}(t) = 1/(1-t)$, and $f_{-1} = 1/(1+t)$, where the latter does not occur in the present calculation. A common abbreviation for the weight vector is to replace occurrences of $k$ zero(s) to the left of $\pm1$ by $\pm(k+1)$. For example, $H_{0,0,1,0,-1}(x) = H_{3,-2}(x)$.
Moreover, all the HPL functions that we have are representable in the form of simple $\Li_n$ functions
of argument $\{x,\bar x, -x/\bar x\}$ with $n=1,2,3,4$. Our result for the color structure $C_F \beta_0$ is in full agreement with~\cite{Melic:2001wb}. Besides, we note that the terms in~\eqref{eq:K_N} proportional to $x$ are reproducible directly from the coefficient function in DIS.

\end{document}